\documentclass[11pt,a4paper]{article}
\usepackage[margin=1in]{geometry}
\usepackage{amsmath,amssymb,bm}
\usepackage{graphicx}
\usepackage[font=small,labelfont=bf]{caption}
\usepackage[expansion=false]{microtype}
\usepackage{xcolor}
\usepackage{booktabs}
\usepackage{array}
\definecolor{link}{RGB}{20,33,61}
\usepackage[colorlinks=true,linkcolor=link,citecolor=link,urlcolor=link]{hyperref}
\usepackage{enumitem}
\setlist{nosep}

\newcommand{\ket}[1]{|#1\rangle}
\newcommand{\bra}[1]{\langle #1|}
\newcommand{\braket}[2]{\langle #1|#2\rangle}
\newcommand{\alw}{\alpha_w}
\newcommand{\abar}{\bar\alpha}
\newcommand{\hb}{\hbar}

\title{Time Running Backwards, and Where It Stops:\\ Superoscillations of Proper Time and the Memory of the Environment}
\author{Yakir Aharonov$^{1,2}$ \and Avshalom C. Elitzur$^{3}$}
\date{\small $^{1}$Institute for Quantum Studies, Chapman University, Orange, CA, USA\\
$^{2}$School of Physics and Astronomy, Tel Aviv University, Tel Aviv, Israel\\
$^{3}$Iyar, The Israel Institute for Advanced Research, Zichron Ya'akov, Israel}

\begin{document}
\maketitle

\begin{abstract}
A quantum system prepared in a superposition of proper times -- a clock whose worldline is superposed, or a clock at rest in a superposed gravitational potential -- evolves, in each branch, by the same dynamics for a different duration. Post-selecting the control that sets the proper time turns the superposition into a superoscillation of the time-dilation factor: the system is found to have evolved for a proper time $\alw t$, where $\alw$ is a weak value that may exceed unity, vanish, or be negative. This is the quantum time-translation machine of Aharonov, Anandan, Popescu and Vaidman. We ask how far back such a machine can run an irreversible process. The test system is a sealed cell -- excited atoms, the radiation field they emit into, and the absorbing walls, carried as one quantum object -- and the witness is projective: atoms that had begun to decay before the run are found, in the post-selected sub-ensemble, more excited than when inserted. The question has an exact answer, expressed through the atoms' survival amplitude alone. With two branches the witness holds only while the decay is younger than the memory time of the environment into which the atoms radiate -- the time after which the emission becomes Markovian and the photon leaves. Beyond it, the post-selection amplifies the minute difference between the contents of the cell in the two branches, a difference set by the full energy spread of the atoms and their field rather than by the linewidth, and the witness fails. Reversing a decay past the memory time requires many branches, an environment with a strictly bounded spectrum, and a post-selection probability that falls exponentially with the number of environmental oscillations to be undone. The obstacle to reversal is therefore not entropy but bandwidth -- the same bandwidth that makes the emission irreversible in the first place. We give four ways to superpose a clock rate, from hollow massive shells to inertial velocities, and discuss what the original proposal, with the shell itself in superposition, would add.
\end{abstract}

\section{Introduction}
\label{sec:intro}

\begin{quote}
\small ``If your theory is found to be against the Second Law of Thermodynamics I can give you no hope; there is nothing for it but to collapse in deepest humiliation.'' -- A. S. Eddington \cite{Eddington1928}
\end{quote}

The scheme presented below does not contradict the second law. It exhibits a sub-ensemble of runs in which a decay -- spontaneous emission -- is found to have run backwards, through the relativity of proper time itself rather than through any dynamical time-reversal operation, and it asks how far back such a reversal can reach. The answer, derived exactly in Section~\ref{sec:memory}, is: as far as the memory of the environment into which the atoms radiate, and beyond it only at a cost that grows exponentially with the depth of the reversal. How this fits Eddington's dictum is the subject of Section~\ref{sec:price}.

Two additions to quantum mechanics, both due to one of us, make the construction possible. The first is \emph{pre- and post-selection}: under suitable pairs of initial and final boundary conditions, a quantum system's intermediate behavior is governed by weak values, which may lie outside the spectrum of the corresponding operator \cite{ABL1964,AAV1988,AharonovRohrlich2005}. The second is \emph{superoscillation}: a superposition of waves whose frequencies are bounded may nevertheless oscillate, locally, faster than any of its components, and the local frequency is a weak value \cite{AAPV1990,Berry1994,BerryPopescu2006,AharonovEtAl2017}. Applied to time evolution, the two together give the quantum time-translation machine \cite{AAPV1990}: a superposition of evolutions, post-selected, acts as a single evolution for a duration outside the superposed range -- longer than the longest branch, or negative. Versions with two superposed evolutions were realized with coupled nuclear spins \cite{Suter1993}, although whether two evolutions capture the machine has been questioned \cite{Vaidman1995} -- a question to which Section~\ref{sec:memory} gives a quantitative answer for irreversible dynamics. For recent work on superoscillations and time translation see \cite{AharonovCohenShushiTBA}.

The gravitational realization proposed in \cite{AAPV1990} -- a hollow massive shell in a superposition of radii, so that the clock rate inside is superposed -- was the starting point of the present work. Its subject here is not the machine but what the machine can be made to reverse, and how one would know. Four things are new.
\begin{enumerate}[label=(\roman*)]
\item \emph{The sealed cell} (Section~\ref{sec:cell}). The test system is not an atom but a small sealed cell containing excited atoms, the radiation field they emit into, and the absorbing walls, carried as one quantum object, superposed, and recombined whole.
\item \emph{The witness} (Section~\ref{sec:atoms}). Freshly excited atoms cannot certify a clock running backwards, because the excited state is a peak of the survival probability in both time directions. Atoms that have already begun to decay can: found more excited than when inserted, they witness a negative effective proper time by a projective measurement that no forward evolution can mimic.
\item \emph{An exact analysis} (Section~\ref{sec:memory}). Everything the machine does to a decaying atom can be expressed through its survival amplitude. With two branches, the witness holds only within the memory time of the environment -- the coherent onset of the emission -- because the post-selection amplifies the which-branch information carried by the full energy spread of the atoms and their field. Beyond it, reversal requires many branches and an exponentially small post-selection probability.
\item \emph{The thermodynamic reading} (Section~\ref{sec:price}). What limits the reversal is not the entropy of an absorber but the bandwidth of the environment, which is also what makes the emission irreversible.
\end{enumerate}
The machine needs one thing from the outside world: a control that puts the cell into a superposition of clock rates. Section~\ref{sec:realizations} gives four, from the original to the simplest -- the hollow shells of \cite{AAPV1990} used classically, heights in the Earth's field, a rotating disk, and inertial velocities -- with their numbers and their leakage channels; the physics of Sections~\ref{sec:cell}--\ref{sec:price} is the same for all. In every one of them the gravitational field, or the velocity, is classical; what is superposed is the cell's worldline. Section~\ref{sec:shell} returns to the version in which the field itself is superposed and states what it would add.

\section{The machine: a superposition of proper times}
\label{sec:machine}

\subsection{Setting}
\label{sec:setting}

Let the internal dynamics of a system be generated by a Hamiltonian $H$ in the system's own rest frame. Let the system's proper time advance, relative to the time $t$ of a fixed external clock, at a rate that depends on a control variable $x$: $d\tau = \alpha(x)\, dt$. The control may be a position in a gravitational potential, a radius on a rotating platform, or a velocity; for now it is a label. Prepare the control in a superposition of $N$ values,
\begin{equation}
\ket{i} = \sum_n a_n \ket{x_n},
\label{eq:pre}
\end{equation}
so that the dilation factor becomes an operator on the control,
\begin{equation}
\hat\alpha = \sum_n \alpha_n \ket{x_n}\bra{x_n}, \qquad \alpha_n = \alpha(x_n),
\label{eq:alphahat}
\end{equation}
and the evolution over external time $t$ is
\begin{equation}
U(t) = \sum_n \ket{x_n}\bra{x_n} \otimes e^{-i\alpha_n H t/\hb}.
\label{eq:U}
\end{equation}
This is a superposition of time translations of the internal system, controlled by $x$ and by nothing else: there is no interaction Hamiltonian between control and system, and in each branch the internal dynamics is the same, run for a different proper time. Everything in this section holds for any control; the controls themselves are the subject of Section~\ref{sec:realizations}.

\subsection{Pre- and post-selection}
\label{sec:tsvf}

The two-state-vector formalism \cite{ABL1964,AAV1988,AharonovRohrlich2005} describes a quantum system between a preparation $\ket{i}$ and a later selection $\ket{f}$. Any observable $\hat A$ then governs a weakly coupled degree of freedom through its weak value,
\begin{equation}
A_w = \frac{\bra{f}\hat A\ket{i}}{\braket{f}{i}},
\label{eq:weak}
\end{equation}
which, when $\braket{f}{i}$ is small, may lie far outside the spectrum of $\hat A$. Anomalous weak values have been measured since 1991 \cite{Ritchie1991}, and Pusey \cite{Pusey2014} showed that they are proofs of contextuality: no non-contextual assignment of values to unmeasured observables reproduces them. We shall need this in Section~\ref{sec:atoms}.

\subsection{Superoscillation}
\label{sec:so}

A function is band-limited if its Fourier components lie within $|k| \le k_{\max}$. Such a function may nevertheless oscillate, over a limited region, faster than $k_{\max}$; the standard example \cite{AAPV1990,Berry1994}
\begin{equation}
f_N(x) = \Big(\cos\frac{x}{N} + i a \sin\frac{x}{N}\Big)^N, \qquad a > 1,
\label{eq:fN}
\end{equation}
has wavenumbers in $[-1,1]$ yet, near $x = 0$, behaves as $e^{iax}$. The local wavenumber $k_{\rm loc}(x) = -i\, d\ln f/dx$ is exactly the weak value of momentum for post-selection at $x$: superoscillation is an anomalous weak value in disguise. Two costs come with it. The amplitude in the superoscillatory region is exponentially small in $N$, and free evolution disperses the superoscillation back into the band \cite{BerryPopescu2006}. Both will return, as the price and the band condition.

\subsection{Post-selection turns the superposition into a superoscillation}
\label{sec:combined}

At the end of the run the control is post-selected in a state $\ket{f} = \sum_n b_n \ket{x_n}$ complementary to the branch label -- for a superposition of positions, a momentum eigenstate; for a superposition of velocities, a position -- so that the post-selection reveals no branch and projects onto a coherent superposition of all of them. The internal system is then acted on by the conditional operator
\begin{equation}
\mathcal{K}_f = \bra{f}U(t)\ket{i} = \braket{f}{i} \sum_n c_n\, e^{-i\alpha_n H t/\hb}, \qquad
c_n = \frac{b_n^* a_n}{\braket{f}{i}}, \qquad \sum_n c_n = 1.
\label{eq:Kf}
\end{equation}
On an internal energy eigenstate of energy $E$, $\mathcal{K}_f$ multiplies by $g(E) = \sum_n c_n e^{-i\alpha_n E t/\hb}$: a function of $E$ that is band-limited in the Fourier sense, its ``wavenumbers'' $\alpha_n t/\hb$ lying in $[\alpha_{\min},\alpha_{\max}]\, t/\hb$. When $\braket{f}{i}$ is small, the $c_n$ are large and alternating in sign, and $g$ superoscillates. Expanding the exponentials,
\begin{equation}
\sum_n c_n e^{-i\alpha_n E t/\hb} = e^{-i\alw E t/\hb}\Big[1 - \tfrac{1}{2}\Big(\tfrac{Et}{\hb}\Big)^2 \sum_n c_n (\alpha_n - \alw)^2 + \dots\Big],
\qquad
\alw = \sum_n c_n \alpha_n = \frac{\bra{f}\hat\alpha\ket{i}}{\braket{f}{i}},
\label{eq:expand}
\end{equation}
the first-order term vanishes exactly when the common exponent is the weak value of the dilation factor. Whenever the higher terms are negligible,
\begin{equation}
\mathcal{K}_f \simeq \braket{f}{i}\; e^{-i\alw H t/\hb}:
\label{eq:machine}
\end{equation}
the post-selected internal system has simply evolved for proper time $\alw t$, and $\alw$, being a weak value, is not confined to the range spanned by the $\alpha_n$. Three regimes follow.
\begin{itemize}
\item $\alw > 1$: \emph{faster than infinity.} No clock at rest in any gravitational potential, and no moving clock, outruns the clock at rest at infinity; for gravity this is the content of the positive-energy theorem \cite{SchoenYau1979,Witten1981}. The sub-ensemble behaves as if sourced by negative mass, the gravitational cousin of negative kinetic energy \cite{APRV1993}.
\item $\alw = 0$: \emph{frozen.} An effective horizon with no black hole; the internal system is returned exactly as inserted.
\item $\alw < 0$: \emph{backwards.} $\mathcal{K}_f \propto e^{+iH|\alw| t/\hb}$: proper time runs the other way. This is the effect announced in the title.
\end{itemize}

The coefficients $c_n$ have a direct meaning. They are the weak values of the projectors onto the branches, $c_n = \langle f|x_n\rangle\langle x_n|i\rangle/\braket{f}{i}$ -- the weak presence of the control in each branch -- and $\alw$ is the average of the clock rates weighted by these presences. The post-selected cell is thus a particle of the three-box kind \cite{AharonovVaidman1991,Aharonov2017}: its presences sum to one, but individually they may be large, and they may be negative. Consider two classical shells of a ton each, of radii one and two meters, with the post-selection chosen so that $\alw = -1$. The cell's presence is then about $+5\times 10^{24}$ in the deeper shell and $-5\times 10^{24}$ in the shallower one. Weighted by its positive presence, the deeper shell slows the clock by four units; weighted by its negative presence, the shallower shell speeds it by two; the net rate is $1 - 4 + 2 = -1$. Time runs backwards because a negative cell sits in the faster shell.

\subsection{Where it holds, and what it costs}
\label{sec:validity}

For two branches with equal weights, $c_{1,2} = (1 \mp K)/2$ where $K$ is the amplification factor, $\alpha_{1,2} = \abar \mp \delta$, and $\alw = \abar - K\delta$. The weak variance in Eq.~\eqref{eq:expand} is $\delta^2(1 - K^2) \simeq -(\abar - \alw)^2$ for $K \gg 1$: it is set not by the actual spread $\delta$ of dilation factors but by the extrapolation distance. The condition for Eq.~\eqref{eq:machine}, which we shall call the \emph{band condition}, is therefore
\begin{equation}
|E - \bar E|\; t\; |\abar - \alw| \ll \hb
\label{eq:band}
\end{equation}
for every energy $E$ carrying appreciable weight in the internal state, $\bar E$ being the mean energy, whose known phase per branch is absorbed into the $b_n$. The name is from superoscillation: the internal energies are the ``frequencies'' of $g(E)$, and the superoscillatory behavior holds only while they cannot resolve the superposed proper times. An internal system able to resolve them falls out of the band and the effect collapses; this is Berry and Popescu's dispersion \cite{BerryPopescu2006} in another guise. The condition concerns the entire energy distribution of the internal state, its tails included, and not only its width; for a decaying atom this is decisive (Section~\ref{sec:memory}).

The price is the probability of the post-selection,
\begin{equation}
P_f \simeq |\braket{f}{i}|^2 = \frac{1}{1 + K^2} \simeq \Big(\frac{\delta}{\abar - \alw}\Big)^2,
\label{eq:price}
\end{equation}
and it is governed entirely by $\delta$, the difference in clock rate between the branches. Deeper reversals require $N > 2$ branches and cost exponentially in $N$. The numbers, for each of the four controls, are in Table~\ref{tab:controls}; all of them are absurd, some less so than others, and we say them aloud because they disarm the obvious objection. This is a gedankenexperiment, and its interest lies in what it shows to be possible in principle and in the conditions under which it is.

\section{The sealed cell}
\label{sec:cell}

The time machine reverses whatever $H$ generates. If $H$ is the Hamiltonian of a spin in a field, the reversal is a rotation, and nothing irreversible has been undone. The interest of the present paper begins when $H$ generates an irreversible process -- spontaneous emission, with the photon absorbed by matter -- and the question is what the test system must be for the reversal to reach it.

Let the test system be a small sealed cell of mass $m$ containing excited atoms, the radiation field inside the cell, and the cell's walls, which absorb. The cell's center of mass is the control $x$ of Eq.~\eqref{eq:pre}: it is put in a superposition of $N$ values of whatever sets its clock rate, held there for a time $t$, recombined, and post-selected (Fig.~\ref{fig:cell} shows the case of classical shells). Three remarks fix what the cell must be.

\emph{One cell, not $N$.} The first idea that comes to mind -- $N$ separate cells, one per branch, with the atoms' wavefunction split among them -- fails, and the reason it fails will organize everything that follows. A photon emitted in cell $n$ remains in cell $n$ when the atoms are extracted and recombined; the decayed component of the atomic state is then labeled by a perfect record of the branch and does not interfere. No measurement by the experimenter is involved: the environment has recorded which branch. The cell must therefore be a single quantum object, carrying atoms, field and walls together, recombined whole. The rule is: \emph{a process can be reversed only if everything it emits is carried by what is recombined.}

\emph{Sealed.} During the run nothing leaves the cell: no photon, no phonon into the support, no heat. Every quantum that escapes is a record of the branch, as above. Emitted photons that reach the walls are absorbed there; the walls are part of the system, and their state is part of $H$'s domain. Section~\ref{sec:walls} states what this requires of them.

\emph{Small, but recombined.} The cell has no lower size limit set by the photon's flight, since the walls absorb and the emission is irreversible in the ordinary sense whatever the cell's size. What sets the scale is that the cell's center of mass must remain coherent across the branches for the duration of the run and be recombined at the end. A gram-scale object in a coherent superposition of two worldlines is far beyond present art; it is six orders of magnitude lighter than a one-ton shell in superposition, and everything below is about what is possible in principle.

One more consequence of sealing will matter in Section~\ref{sec:walls}. Because nothing leaves, the total internal energy of the cell is the same in every component of its state: ``atom excited, wall cold'' and ``atom in its ground state, wall carrying one photon's energy'' have the same mean energy. The rest-energy phase that different internal energies would acquire in different branches -- the phase of Zych et al.\ \cite{Zych2011}, which dephases a clock in superposition in Pikovski et al.\ \cite{Pikovski2015} -- is here a common factor and cancels.

\section{The witness}
\label{sec:atoms}

\subsection{What freshly excited atoms can and cannot certify}
\label{sec:fresh}

Write the free evolution of an excited atom in the vacuum,
\begin{equation}
e^{-iH\tau/\hb}\ket{e,0} = A(\tau)\ket{e,0} + \sum_k \beta_k(\tau)\ket{g,k}, \qquad |A(\tau)| \simeq e^{-\Gamma|\tau|/2}.
\label{eq:decay}
\end{equation}
Hermiticity of $H$ gives $A(-\tau) = A(\tau)^*$: the survival amplitude is even in $\tau$. The excited state is a \emph{peak} of the survival probability, and one descends from it in either time direction. An excited atom evolved backwards de-excites into a photon converging upon it -- mirror decay.

This settles what fresh atoms can do. Every forward branch, at any rate $\alpha_n \in [\alpha_{\min},\alpha_{\max}]$, leaves an excited fraction between $e^{-\Gamma\alpha_{\max}t}$ and $e^{-\Gamma\alpha_{\min}t}$, and so does every mixture of branches. A post-selected $\alw > \alpha_{\max}$ gives $e^{-\Gamma\alw t}$, below the floor:
\begin{equation}
p_{\rm exc}(f) < e^{-\Gamma\alpha_{\max}t},
\label{eq:ineq-fast}
\end{equation}
\emph{atoms more decayed than the fastest branch allows.} In the ideal limit of Eq.~\eqref{eq:machine}, fresh atoms would therefore certify the first anomaly, a clock faster than any branch and, for $\alw > 1$, faster than infinity; but the which-branch term of Section~\ref{sec:memory} also lowers the excited fraction, and this witness cannot be used without separating the two. Fresh atoms cannot certify the second anomaly at all. A negative $\alw$ gives $e^{-\Gamma|\alw|t}$, which is above the floor but never above the ceiling $e^{-\Gamma\alpha_{\min}t}$ -- survivors in excess of the slowest branch would witness at most a frozen time, $\alw = 0$, and even those are ordinary, since a fraction $e^{-\Gamma t}$ survives with no post-selection at all. Finding excited atoms proves nothing about reversal.

\subsection{The remedy: give the inserted state an arrow}
\label{sec:predecayed}

\begin{figure}[t]
\centering
\includegraphics[width=\textwidth]{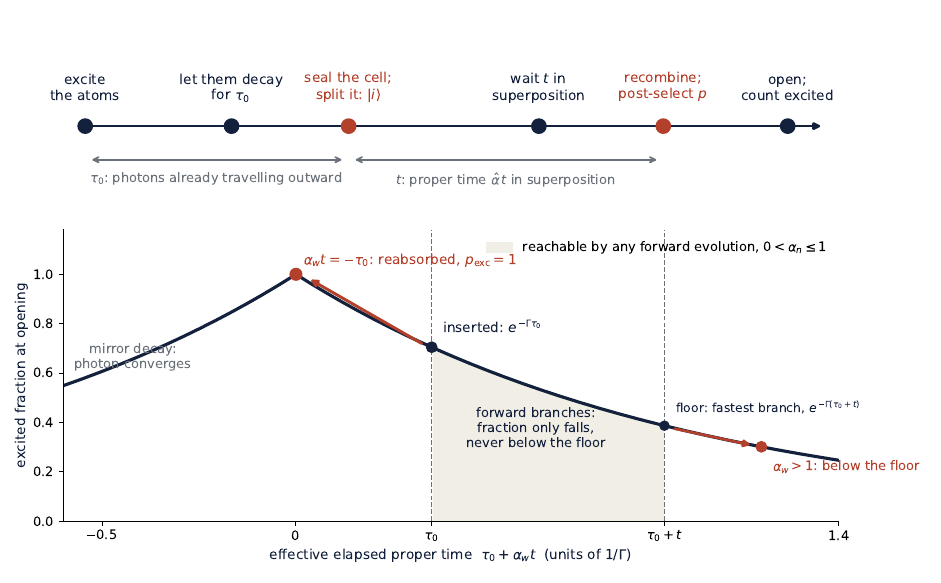}
\caption{Top: the protocol. Excite the atoms; let them decay for $\tau_0$; seal the cell and split its control; wait $t$; recombine and post-select; open and count. Bottom: in the ideal limit of Eq.~\eqref{eq:machine}, the excited fraction at opening against the effective elapsed proper time $\tau_0 + \alw t$. Every forward evolution stays in the shaded region, below the inserted value $e^{-\Gamma\tau_0}$ and above the floor set by the fastest branch; frozen time returns the inserted value; a negative $\alw$ raises the fraction, up to $1$ at $\alw t = -\tau_0$, beyond which the atoms decay again by mirror emission; an $\alw$ above the fastest branch lowers it below the floor.}
\label{fig:protocol}
\end{figure}

Insert atoms that have already begun to decay. Excite them; let them evolve freely for a time $\tau_0$, so that the emission has begun (Section~\ref{sec:memory} shows how short $\tau_0$ must be); only then seal the cell, split its control into $\ket{i}$, and wait a time $t$ (Fig.~\ref{fig:protocol}). The inserted internal state $\ket{\psi(\tau_0)} = e^{-iH\tau_0/\hb}\ket{e,0}$ has an arrow built in. Now Eq.~\eqref{eq:machine} gives
\begin{equation}
\mathcal{K}_f\ket{\psi(\tau_0)} \propto \ket{\psi(\tau_0 + \alw t)}, \qquad
p_{\rm exc}(f) = e^{-\Gamma|\tau_0 + \alw t|},
\label{eq:witness}
\end{equation}
and the three regimes separate completely:
\begin{itemize}
\item Forward, in every branch and at every rate $\alpha_n > 0$: the photon continues outward; the excited fraction lies \emph{below} $e^{-\Gamma\tau_0}$, in every branch and every mixture of branches.
\item Frozen, $\alw = 0$: the state is returned as inserted; the excited fraction is exactly $e^{-\Gamma\tau_0}$.
\item Backward, $\alw t = -\tau_0$: $\mathcal{K}_f$ applies $e^{+iH\tau_0/\hb}$ to $e^{-iH\tau_0/\hb}\ket{e,0}$; the outgoing photon returns and is reabsorbed; the excited fraction is $1$.
\end{itemize}
The operational statement is an inequality:
\begin{equation}
p_{\rm exc}(f) > e^{-\Gamma\tau_0}:
\label{eq:ineq}
\end{equation}
\emph{atoms more excited than when inserted.} Nothing that evolves forward in any branch, or in any mixture of branches, can leave the atoms above the inserted value. No assignment of forward proper times to the branches accounts for the excess; it is an interference term between branches, and in Pusey's sense \cite{Pusey2014} a contextuality witness. Nor can a disturbance account for it: any run in which something -- the measurement, the environment, or anything else -- has made the branch definite is a forward run, and obeys the inequality. Disturbance can only reduce the violation; it cannot produce it. The readout is projective: each outcome, ``excited'' or ``ground'', is a legal eigenvalue, and the anomaly lies in its frequency -- the same logic as in the quantum pigeonhole and the interaction-free effects between distant atoms. This is what distinguishes the present proposal from a weak measurement of an anomalous time \cite{Angulo2024}: the sign of the effective proper time is read from a strong measurement on a single run, and the statistics of such runs are what no forward evolution can produce. Equation~\eqref{eq:witness} is the ideal limit; Section~\ref{sec:memory} computes the excited fraction exactly and finds where the inequality survives.

In the ideal limit, the photon itself corroborates the reading. At insertion it is in a shell of radius $c\tau_0$ about the atoms; forward it is farther out, frozen it is where it was, backward it is gone. And the symmetry has a further face: post-selecting $\alw t = -2\tau_0$ yields atoms in the ground state with a photon converging upon them -- the advanced wave -- showing that the machine passes through the excited state and continues.

\section{How far back: the memory of the environment}
\label{sec:memory}

\subsection{An exact expression}
The machine acts on the cell's contents only through the operators $e^{-i\alpha_n Ht/\hb}$, which commute with $H$. Everything it does to a decaying atom can therefore be expressed through one function, the survival amplitude $A(\tau) = \bra{e,0}e^{-iH\tau/\hb}\ket{e,0}$ of Eq.~\eqref{eq:decay}. For atoms inserted after a free decay of duration $\tau_0$, the excited fraction in the post-selected runs is, exactly,
\begin{equation}
p_{\rm exc}(f) = \frac{\big|\sum_n c_n\, A(\tau_0 + \alpha_n t)\big|^2}{\sum_{n,m} c_n c_m^*\, A\big((\alpha_n - \alpha_m)\,t\big)} .
\label{eq:exact}
\end{equation}
The numerator is the amplitude to find the atom excited. The denominator is the norm of the post-selected internal state: the overlaps between the cell's contents in different branches, weighted by the large coefficients $c_n$. In the ideal limit of Eq.~\eqref{eq:machine} it equals one; in general it measures how well the cell's contents tell the branches apart -- the which-branch information of Section~\ref{sec:whichbranch}.

\subsection{Two branches}
For two branches with $K \gg 1$, the numerator of~\eqref{eq:exact} is the linear extrapolation $A(T) - (\abar - \alw)\,t\,A'(T)$, with $T = \tau_0 + \abar t$, which for an exponential decay reproduces $e^{-\Gamma(\tau_0 + \alw t)/2}$ to first order: this is the effect. The denominator is
\begin{equation}
1 + \frac{K^2 - 1}{2}\,\big[1 - \mathrm{Re}\,A(2\delta t)\big] \;\simeq\; 1 + \Big(\frac{\Delta H\,(\abar - \alw)\,t}{\hb}\Big)^2 ,
\label{eq:norm}
\end{equation}
where $\Delta H$ is the energy spread of the inserted state; the approximation holds because $2\delta t$ is far shorter than every time scale of $A$. The amplification $K^2$ acts on the minute difference between the branches, and that difference is set not by the linewidth $\hb\Gamma$ but by the full spread $\Delta H$. For a decaying atom $\Delta H \gg \hb\Gamma$: the spread is dominated by the far wings of the line, the field modes far from resonance with which the atom exchanges energy virtually. If the line were exactly Lorentzian, $\Delta H$ would be infinite, the bracket in~\eqref{eq:norm} would be linear rather than quadratic in $\delta t$, and the denominator, growing in proportion to $K$, would destroy the effect at any duration.

\subsection{The memory time}
A realistic line is Lorentzian near resonance and cut off at the bandwidth $\Lambda$ (an angular frequency) of the environment into which the atom radiates. Its decay has two stages \cite{MisraSudarshan1977,FondaGhirardiRimini1978}. For times short compared with the memory time $\tau_c = 1/\Lambda$ of the environment, the excited population falls quadratically, and the emission is still coherent with the atom; after $\tau_c$ it falls exponentially at the rate $\Gamma$, the decay is Markovian, and the photon leaves for good. For a Lorentzian line multiplied by a Lorentzian cutoff,
\begin{equation}
A(\tau) = \frac{\Lambda\, e^{-\Gamma|\tau|/2} - \tfrac{\Gamma}{2}\, e^{-\Lambda|\tau|}}{\Lambda - \tfrac{\Gamma}{2}}, \qquad \Delta H^2 = \tfrac{1}{2}\,\hb^2\Gamma\Lambda .
\label{eq:model}
\end{equation}

\begin{figure}[t]
\centering
\includegraphics[width=0.92\textwidth]{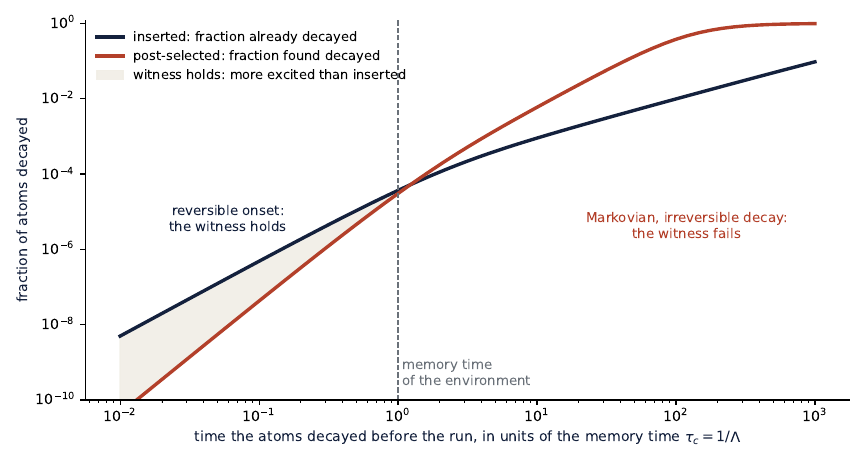}
\caption{The exact excited fraction, Eq.~\eqref{eq:exact}, for the amplitude~\eqref{eq:model} with $\Lambda = 10^4\,\Gamma$, two branches, $t = \tau_0/10$ and full reversal ($\alw t = -\tau_0$), shown as the fraction of atoms decayed. Navy: when inserted. Red: as found in the post-selected runs. The witness -- fewer atoms decayed than when inserted -- holds while the decay is younger than the memory time $\tau_c = 1/\Lambda$ of the environment, and fails once the decay has become Markovian.}
\label{fig:memory}
\end{figure}

The witness requires the gain in the numerator of~\eqref{eq:exact} to exceed the loss in its denominator, roughly $(\Delta H\,(\tau_0 + t)/\hb)^2 < 1 - |A(\tau_0)|^2$. In the Markovian stage the right-hand side is $\Gamma\tau_0$ and the left-hand side $\Gamma\Lambda(\tau_0 + t)^2/2$: the inequality fails as soon as $\tau_0$ exceeds about $\tau_c$. In the coherent stage both sides are of order $(\Delta H\,\tau_0/\hb)^2$, the error enters only at higher order, and the witness holds. An exact evaluation of~\eqref{eq:exact} with the amplitude~\eqref{eq:model} confirms this (Fig.~\ref{fig:memory}). With $t = \tau_0/10$, the witness holds up to $\tau_0 \simeq 1.2\,\tau_c$, independently of $\Gamma/\Lambda$ over three decades of $\Lambda$, and fails increasingly beyond; at $\tau_0 = 100\,\tau_c$ the post-selected runs are dominated by decayed atoms.

For atoms radiating into free space the memory time is of the order of the optical period, about a femtosecond. In that time the emitted field has not separated from the atom by more than a wavelength, and no wall is involved. With two branches, the machine reverses the coherent onset of the emission, not the emission itself.

\subsection{Many branches}
Beyond $\tau_c$, the post-selected combination must reproduce $e^{+iH\tau_0/\hb}$ over the whole spectrum of the cell's contents, across which it oscillates about $\Lambda\tau_0$ times. A superoscillation with that many oscillations requires at least as many branches, and its coefficients -- hence the inverse of the post-selection probability -- grow exponentially with that number \cite{Berry1994,FerreiraKempf2006}. The spectrum must, moreover, be strictly bounded, since any weight outside it is amplified by the same coefficients. Reversing a nanosecond of optical decay would require of order $10^6$ branches. Such a reversal is allowed in principle; its price is the subject of Section~\ref{sec:price}.

\section{Four ways to superpose a clock rate}
\label{sec:realizations}

Nothing in Sections~\ref{sec:machine}--\ref{sec:memory} depends on what the control is. What we want it to do is two things: give the branches different clock rates, and give them nothing else. Where it gives them something else, that something must be either common to all branches or a known phase that the post-selection can absorb. Any difference between branches that is neither we call a leakage channel. It is a degree of freedom whose final state depends on the branch; it reduces the overlap of the branches, and with it the interference the witness relies on. We describe four controls, from the original to the simplest, and show at each step what is gained in price and what is paid in leakage. Table~\ref{tab:controls} summarizes them.

\begin{figure}[t]
\centering
\includegraphics[width=\textwidth]{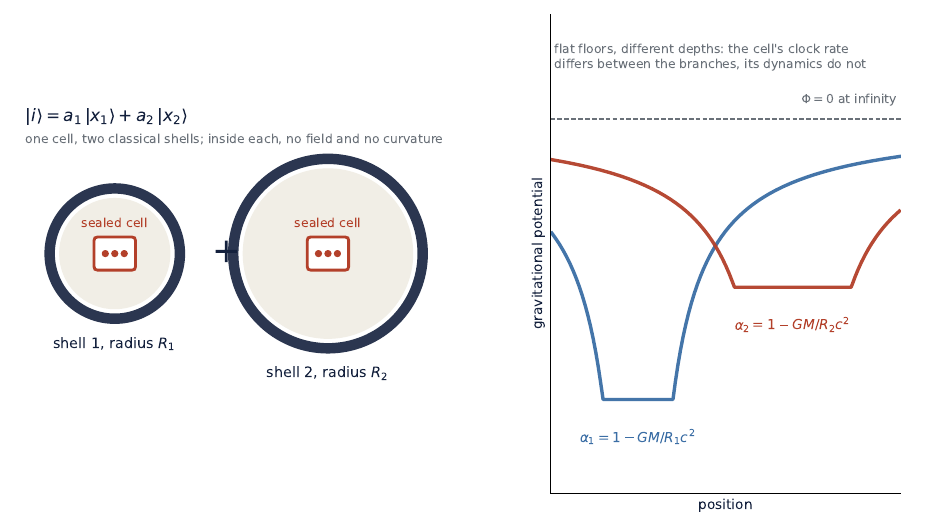}
\caption{Left: one sealed cell in a superposition of two positions, each inside a classical hollow shell. The atoms, the field and the absorbing walls are all inside the cell; the shells are not superposed. Right: the potential along a line through both shells. Inside each shell the potential is flat and the field vanishes; the two floors differ in depth, hence in clock rate $\alpha$. In each branch the internal dynamics is the same and only the proper time elapsed differs. The other controls of this section replace the shells by heights, radii on a disk, or velocities.}
\label{fig:cell}
\end{figure}

\emph{Hollow massive shells, used classically.} Newton's shell theorem states that a uniform spherical shell exerts no force on a body inside it. General relativity preserves this result and adds to it: by Birkhoff's theorem \cite{Birkhoff1923} the spacetime outside a static shell of mass $M$ and radius $R$ is Schwarzschild, and inside it is exactly flat, with time running at
\begin{equation}
\alpha_R = \sqrt{1 - \frac{2GM}{Rc^2}} \simeq 1 - \frac{GM}{Rc^2}
\label{eq:shell}
\end{equation}
relative to a distant observer -- no force, no curvature, no local experiment that detects anything, yet clocks slow by a factor set by the potential alone. This is the control of \cite{AAPV1990}, with one change: there the shell was in a superposition of radii, here it is not. Two classical shells of different radii have interiors that differ in nothing but the constant $\alpha_R$, and the cell is put in a superposition of positions inside them (Fig.~\ref{fig:cell}); where it sits within a shell is immaterial, since the potential is uniform throughout the interior, and only the difference in depth between the shells matters. Then $\delta = \tfrac{1}{2}|GM_1/R_1 - GM_2/R_2|/c^2$, about $4 \times 10^{-25}$ for laboratory shells of a ton and a meter. This is the purest control -- no force, no field, no leakage channel of any kind, the branches differing in one number and in nothing else -- and the most expensive.

\emph{Heights in the Earth's field.} Drop the shells and put the cell in a superposition of two heights $h_1$ and $h_2$, both held against gravity. Then
\begin{equation}
\delta \simeq \frac{g\,(h_2 - h_1)}{2c^2},
\label{eq:earth}
\end{equation}
about $5 \times 10^{-17}$ for a meter -- eight orders of magnitude better than the shells, at no cost: the field is common to both branches and drops out, so that the branches again differ in nothing but the constant clock rate. There is no leakage channel beyond the requirement that the center of mass stay coherent across a meter of height.

\emph{A rotating disk.} Replace the Earth's field by a centrifugal one. Mount the cell on a platform rotating at angular velocity $\omega$, with its center of mass in a superposition of two radii $r_1$ and $r_2$. The clock rate at radius $r$ is $\sqrt{1 - \omega^2 r^2/c^2}$, so that
\begin{equation}
\delta \simeq \frac{\omega^2 (r_2^2 - r_1^2)}{4c^2},
\label{eq:disk}
\end{equation}
about $10^{-12}$ for a rim speed of $500$ meters per second and a cell a few centimeters from the axis in one branch and near the rim in the other: four more orders of magnitude, on a platform of centimeters. In the rotating frame this is a superposition of two depths in the centrifugal potential, by the equivalence principle a superposition of gravitational potentials; in the laboratory frame it is a superposition of two speeds, kept a few centimeters apart. Two things are common to the branches: the gradient of the centrifugal field, $\omega^2$, so that tidal effects across the cell do not record the branch; and the Sagnac phase of the light inside the cell, which depends on $\omega$ and on the enclosed area only. Here, for the first time, a leakage channel appears. The uniform part of the centrifugal field, $\omega^2 r$, differs between the branches, and a cell supported against a different uniform field is in a different internal state -- its walls strained differently, the atoms trapped in them displaced differently. The overlap of two ground states of a macroscopic body under slightly different stress is a computable quantity, and for a small stiff cell at moderate $\omega$ it is presumably close to unity, but it must be estimated.

\emph{Inertial velocities.} Finally, drop the platform. Give the cell's center of mass two momentum components, so that it moves at speeds $v_1$ and $v_2$ relative to the laboratory, and recombine the two wave packets at the end of the run; the post-selection is on position, at the output where the branches overlap. Proper time in branch $n$ is $t/\gamma_n$, so that
\begin{equation}
\delta \simeq \frac{v_2^2 - v_1^2}{4c^2},
\label{eq:inertial}
\end{equation}
about $3 \times 10^{-12}$ for speeds of a kilometer per second. This is the twin paradox in superposition with post-selection, and as an interferometer it exists: it is the clock interferometry of Zych et al.\ \cite{Zych2011} in its kinematic form, and the single-electron realization proposed by Bushev et al.\ \cite{Bushev2016}. The geometry must be that of the twins, not of a symmetric Mach--Zehnder interferometer: in the latter the mirror exchanges the velocities of the two arms halfway, each arm is fast for half the run and slow for the other half, and the two accumulate equal proper time. Here one arm stays at rest throughout while the other goes out at speed $v$, is reflected, and returns; the two meet at the same place with different velocities, a final splitter mixes them, and the two output ports are the two superpositions of the arm states -- the dark one being the post-selection $\ket{f}$ nearly orthogonal to $\ket{i}$. During free flight the branches differ in nothing but $\gamma$: by Lorentz invariance the internal physics in the cell's rest frame is identical in both, and no field acts. The leakage channel is the turnaround. Bringing two branches of different speed to a common final velocity requires accelerations that differ between branches, and an accelerated cell is not in the same internal state as an unaccelerated one; the difference is confined to the turnaround, and it must be shown negligible, or made common to both branches by decelerating each with the same profile and letting the difference sit in the center-of-mass phase.

\begin{table}[t]
\centering
\small
\begin{tabular}{@{}p{3.1cm}p{3.6cm}p{3.5cm}p{4.4cm}@{}}
\toprule
Control & What differs between branches & $\delta$ for stated parameters & Leakage channel beyond the clock rate \\
\midrule
Classical shells & A constant clock rate; no field in either branch & $4\times 10^{-25}$ for a ton at a meter & None \\
Heights on Earth & Depth in a common uniform field & $5\times 10^{-17}$ per meter & None; the field is common \\
Rotating disk & Depth in the centrifugal potential & $10^{-12}$ at $500$~m/s rim speed & The uniform centrifugal field differs; strain state of the cell \\
Inertial velocities & Speed; no field in either branch & $3\times 10^{-12}$ at $1$~km/s & The turnaround: branch-dependent acceleration \\
\midrule
\multicolumn{4}{@{}p{15.2cm}@{}}{Post-selection probability for $\alw = 0$, Eq.~\eqref{eq:price}: $P_f \simeq \delta^2$, i.e.\ one run in $10^{48}$, $3\times 10^{32}$, $2\times 10^{24}$ and $10^{23}$ respectively.} \\
\bottomrule
\end{tabular}
\caption{Four controls for a superposition of clock rates, from the original to the simplest. In all four the gravitational field, where there is one, or the velocity, is classical; what is superposed is the cell's worldline. The physics of Sections~\ref{sec:cell}--\ref{sec:price} is common to all.}
\label{tab:controls}
\end{table}

Going down Table~\ref{tab:controls}, the price falls by twenty-five orders of magnitude and the purity falls with it. The shell is where the machine of \cite{AAPV1990} was first written down, and it remains the only control in which the branches differ in a single number and nothing acts on the cell at all; the kinematic controls buy their probability at the cost of a leakage channel that must be estimated. None of the four says anything about the gravitational field: each is quantum mechanics on a fixed background, curved or flat. Section~\ref{sec:shell} is about the fifth control, which does.

\subsection{The clock as its own which-branch detector}
\label{sec:whichbranch}

One question applies to all four controls and should be faced before the conditions of Section~\ref{sec:conditions}. The cell's contents are a clock. A clock that has run for different proper times in the two branches is in different states, and its reading is which-branch information: if the atoms could tell how long they had evolved, they would tell which branch they had been in, and the interference the post-selection relies on would be gone. This is the observation of Zych et al.\ \cite{Zych2011}: the visibility of a clock interferometer equals the overlap of the clock states in the two arms, and a clock sharp enough to resolve the proper-time difference destroys the fringes.

We do not prevent this; we work where it cannot happen. The band condition \eqref{eq:band} is precisely the requirement that the internal energy spread, multiplied by the proper-time difference between the branches, be small compared with $\hb$ -- that is, that the clock be too coarse to resolve the difference. The overlap of the states of one atom and its field in the two branches is $A(2\delta t)$, whose deficit from unity, $\simeq 2(\Delta H\,\delta t/\hb)^2$, is set by the full energy spread $\Delta H$ rather than by the linewidth (Section~\ref{sec:memory}); and this loss of visibility, raised to the $N$th power for $N$ atoms, is what sets the threshold on the number of atoms in Section~\ref{sec:number}. Zych's visibility and our atom bound are the same quantity.

That raises the question the reader will have asked: if the clock cannot resolve the superposed proper times, how can the post-selection extract from it a proper time outside their range? The answer is superoscillation. In a band-limited function, too, no local measurement resolves the neighboring frequencies of the band, yet in the region where the amplitude is exponentially small the local frequency exceeds every component. The dark port is that region. There the small difference between the branches, which the clock does not see, is amplified by the factor $K$ of Section~\ref{sec:validity}, and the price of the amplification is the probability $\delta^2$ of Eq.~\eqref{eq:price}. The effect does not exist in spite of the clock's inability to distinguish the branches. It exists because of it.

The same requirement applies to everything else the cell carries. The center of mass must return to the same position and velocity distribution in both branches, or it records the branch itself; and the known phases between the branches -- the kinetic phase of the moving arm, and the rest-energy phase $mc^2\,\Delta\tau/\hb$, which for a gram-scale cell and a proper-time difference of $10^{-20}$~s is of order $10^{28}$ radians -- are absorbed into the choice of which output port is dark. They are known, and that is all the formalism needs; their stability is one more reason that everything here is a gedankenexperiment.

\section{Conditions}
\label{sec:conditions}

\subsection{Duration}
With two branches the duration condition is the memory-time condition of Section~\ref{sec:memory}: the atoms may have decayed before the run, and the run may last, only about as long as the memory time of the environment into which they radiate,
\begin{equation}
\tau_0 + t \lesssim \tau_c ,
\label{eq:duration}
\end{equation}
which for atoms radiating into free space is of the order of a femtosecond. The certifiable effect is a recovery of the excited fraction from the small value lost in the coherent onset of the emission back to one.

\subsection{The walls}
\label{sec:walls}

Within the memory time the emitted field has not separated from the atoms, let alone reached a wall, and the walls play no part in the two-branch effect. They enter only in a reversal deep enough to require many branches, and there what was said of the atom and its field holds for everything the cell contains. The conditional operator~\eqref{eq:machine} acts on $H$, the Hamiltonian of atoms, field and walls together. Where its band condition holds for the whole of it, the reversal inverts the complete unitary evolution -- entanglement between the photons and the modes of the wall included -- without sorting anything: a Loschmidt configuration reached by a boundary condition, the echo without the echo pulse.

What the walls must satisfy is therefore a spectral condition, not a thermodynamic one. A wall at temperature $T$ with heat capacity $C$ is a mixture with an energy spread of order $\sqrt{k_B T^2 C}$, for a gram of metal about $10^{-10}$~J, far outside any band. But the obstacle is the spread, not the entropy: a pure state with the same spread would fail equally, and a closed unitary evolution is reversible whatever its entropy. Sealing remains necessary (Section~\ref{sec:cell}), because whatever leaves the cell is a record of the branch.

\subsection{How many atoms}
\label{sec:number}

The atoms are the pointer that reads the control, as Section~\ref{sec:whichbranch} explained, and $N$ atoms read it $N$ times harder. For one atom the deficit of the overlap between branches is $\simeq 2(\Delta H\,\delta t/\hb)^2$; for $N$ atoms it is multiplied by $N$; and the amplification $K^2 = (\abar - \alw)^2/\delta^2$ undoes the smallness of $\delta$. The norm in the denominator of~\eqref{eq:exact} becomes
\begin{equation}
1 + N\Big(\frac{\Delta H\,(\abar - \alw)\,t}{\hb}\Big)^2, \qquad\text{so that}\qquad N \ll \Big(\frac{\hb}{\Delta H\,(\abar - \alw)\,t}\Big)^2 .
\label{eq:N}
\end{equation}
Below this threshold the success probability is independent of $N$ and every atom in a successful run shows the anomaly; within the memory time the threshold is large, of order $\Lambda/\Gamma$ for the line~\eqref{eq:model}. Above it the anomaly does not fade; it vanishes, and the post-selected populations revert to the ordinary forward value. The bound is informational, not gravitational: the gravitational field of the atoms is irrelevant at any conceivable $N$; what limits $N$ is which-branch information. Collective (Dicke) emission \cite{Dicke1954} is allowed, with the collective rate replacing $\Gamma$.

The independence of $P_f$ from $N$ below the threshold deserves a remark, because it runs against the experience of weak measurements, where every additional probe costs. Here the atoms are not probes coupled weakly to the control at intermediate times; they are the system itself, carried through the run in superposition and read strongly only after the post-selection has succeeded. This is the structure of the quantum shutter \cite{AharonovVaidman2003} and of the disappearing-particle experiments \cite{Aharonov2017,ElitzurCohen2018}: the probe remains in superposition only because the post-selection has already been obtained. Below the threshold, one atom or a hundred are read at the same price.

\section{The thermodynamic price}
\label{sec:price}

Four statements account for Eddington.

\emph{Hindsight.} The control is measured first, the atoms only afterwards, and the reversal is certified only once the outcome of the post-selection is known. This is the condition under which every quantum anomaly is seen: the two-slit particle is known to have passed through both slits only after it has shown interference, and two entangled particles are known to have been correlated only after both results are brought together. Were any of these knowable in real time, locality or causality would fail; here, were the reversal knowable before the post-selection, the future could be known and acted upon, which is what pre- and post-selection, properly understood, never allows \cite{AharonovCohenShushi2016}. The technical content is that the reduced state of the atoms, averaged over the outcomes of the post-selection, is independent of anything done to the control. Post-selection sorts runs; it changes nothing on average. Over the whole ensemble exactly the expected number of atoms decayed, and exactly the expected energy was deposited in the walls. Nothing is resurrected.

\emph{A bound on the sub-ensemble.} By the Cauchy--Schwarz inequality,
\begin{equation}
P(f \wedge \text{excited}) \le \sum_n |a_n|^2\, |A(\tau_0 + \alpha_n t)|^2 = P(\text{excited}):
\label{eq:CS}
\end{equation}
the runs in which the atoms are ``more excited than inserted'' are never more numerous than the ordinary survivors. The conditional probability \eqref{eq:witness} is large only because $P_f$ is small.

\emph{The record.} The post-selection outcome is a macroscopic record. Conditioning on it lowers the entropy of the atoms and walls at the cost of the record and of the discarded runs, exactly as for Maxwell's demon. The second law constrains ensembles and averages; it does not constrain a sub-ensemble selected by a final boundary condition.

\emph{The environment.} It is tempting to say that reversal through an absorber requires an absorber without entropy. Section~\ref{sec:memory} shows that this is not quite the point. A closed unitary evolution is reversible whatever its entropy; what limits the machine is the spectral width of what it must reverse, through the which-branch information that width carries. That width is also what makes the emission irreversible: a broad environment forgets quickly, and it is past its memory time that the photon leaves for good. The second law does not forbid the post-selected reversal; it sets its price, which grows exponentially with the number of environmental oscillations to be undone.

There is a compact way to put this. Entropy increases, as always, along the time of the clock that runs; the machine can make that time run backwards only for as long as the cell's contents have not yet produced any.

\section{Interpretation}
\label{sec:interp}

Two readings fit the same facts. In the two-state-vector reading, the atoms began to emit and then, as effective proper time turned around, reabsorbed what they had begun to emit; the operational content is that the weak value of the photon number at intermediate times rises and returns to zero. In the one-vector, interference reading, nothing propagates backwards: a sum of forward-evolved branches equals, on the band-limited subspace, an earlier state, just as a sum of redshifted waves may locally behave as a blueshifted one.

A natural question is what one would have found by opening the cell early. The answer is partly decayed atoms, and it is right to say so. What one may not do is use it as in classical physics -- infer that in the runs that were not opened the atoms \emph{were} decayed at that moment, and that the later finding is therefore paradoxical. This is the counterfactual of Bell's theorem, and it plays here the role it plays there. Bell's inequality is derived by assuming that unmeasured quantities had values, and its violation shows that they did not; Pusey's theorem \cite{Pusey2014} does the same for anomalous weak values, which are exactly what the post-selected sub-ensemble exhibits. The counterfactual may be stated; the value it would have returned may not be assigned to the runs in which it was not performed. The witness inequality~\eqref{eq:ineq} is independent of the reading.

The negative proper time belongs to a family. In each of its members a projector -- the presence of a system in some region or state -- acquires a negative weak value, and a positive quantity multiplied by that presence turns negative with it. Multiplied by mass, it gives the negative-mass particles of the time-symmetric ontology \cite{Waegell2023} and the counter-particles of \cite{AharonovCarmi2018}; multiplied by the time an atom remains excited, it gives the negative excitation time measured by Steinberg's group \cite{Angulo2024}; multiplied by a clock rate, it gives the negative proper time of the present paper, in which the presence of the cell in the faster branch is negative (Section~\ref{sec:combined}). The family has a second face, because time and energy are conjugate: a clock that evolves under $\alw H$ with $\alw < 0$ evolves as if its internal energies were negative, so that the post-selected cell is, in the language of \cite{Waegell2023}, a counter-particle in energy. Two things distinguish the present member. Proper time scales the entire internal Hamiltonian at once, so that in principle the reversal reaches everything inside the cell -- atoms, field and walls -- rather than a single degree of freedom, at the price computed in Section~\ref{sec:memory}; and the sign is certified not by a weak probe but by projective readouts and inequalities that every forward evolution, with or without disturbance, satisfies.

\section{The fifth control: the shell in superposition}
\label{sec:shell}

We have focused on the superoscillation of time dilation that gives a negative, or a super-unitary, effective proper time, and for that purpose we have examined the alternatives to the shell. The shell, however, raises questions about spacetime that the alternatives do not, and it deserves a discussion of its own.

Each of the four controls of Section~\ref{sec:realizations} keeps the gravitational field classical. The cell is superposed among fixed shells, at fixed heights, on a fixed disk, or between fixed velocities; the geometry is one spacetime, and what is superposed is the cell's worldline within it. Superposed proper time of this kind is the subject of clock interferometry \cite{Zych2011,Pikovski2015,Bushev2016}, and the present proposal is, in that language, a clock interferometer with post-selection, in which the clock is an irreversible process and the readout is projective. It says nothing about gravity as such.

The original proposal \cite{AAPV1990} superposes the source. A hollow shell in a superposition of radii $R_1, R_2$ produces, on the assumption that the field of a superposed mass is itself superposed, two flat interiors differing in $\alpha_R$, with the cell at one place inside. Everything in Sections~\ref{sec:machine}--\ref{sec:interp} carries over unchanged: Eqs.~\eqref{eq:U}--\eqref{eq:N} are the same with $\hat\alpha$ an operator on the shell's radius and the post-selection a measurement of the shell's conjugate variable. (An earlier version of this paper used two parallel plates, one of them in a superposition of positions, as a more tractable source than a shell of variable radius; the interior of the plates is likewise field-free.) The effect certified by the witness is the same.

What the superposed source adds is a statement about the field. If the effect is observed with a superposed shell, the gravitational field has been shown to enter a superposition and to act coherently on a clock -- the question addressed by proposals to witness gravitationally induced entanglement between masses \cite{Bose2017,MarlettoVedral2017}, and one on which mean-field semiclassical gravity \cite{Rosenfeld1963,PageGeilker1981} gives the opposite prediction: a mean field, no superposition of clock rates, no effect. (Classical-gravity models that violate the locality assumptions of the no-go arguments may behave otherwise \cite{TrilloNavascues2024,DiBiagio2025}.) A force version of the same idea -- a single superposed source mass whose post-selection makes gravity effectively repulsive on a probe -- has recently been proposed as a witness of quantum gravity \cite{Saldanha2026}. The superposed shell is thus also a test of whether gravity is quantum, with spontaneous emission rather than an interference phase as the witness. Its price is the one the four controls were chosen to avoid: a body of tons in a coherent superposition of configurations. Because the interior is exactly flat in every branch, the problem of comparing superposed geometries \cite{Penrose1996} does not arise even here; the branches differ in a number, not in a shape, and in the language of quantum reference frames \cite{Giacomini2021,Kabel2023} the two interiors are one geometry seen by two clocks.

We regard the five controls as one proposal at two levels. The four classical ones show that quantum mechanics, on a fixed background, already permits in a post-selected sub-ensemble the reversal of a decay, within the memory of its environment, by a superposition of proper times alone. The superposed shell would show, in addition, that the geometry itself can be the superposed control.

\section{Summary}
\label{sec:summary}

In a post-selected sub-ensemble, the proper time of a sealed system can freeze or run backwards, and the reversal can be certified by a strong measurement that no forward evolution can mimic: atoms found more excited than when inserted. The exact analysis, expressed through the survival amplitude alone, shows how far back the reversal reaches. With two branches it reaches through the memory time of the environment into which the atoms radiate -- the coherent onset of the emission -- and not beyond, because there the post-selection amplifies the which-branch information carried by the full energy spread of the atoms and their field. Past the memory time, reversal requires many branches, a strictly band-limited environment, and an exponentially small post-selection probability. The obstacle is bandwidth, not entropy, and it is the same bandwidth that makes the emission irreversible: the arrow of time enters exactly where the machine stops. Any control that superposes the cell's clock rate will do -- shells, heights, a rotating disk, or velocities -- at prices that differ by twenty-five orders of magnitude; and the superposed shell is where the same construction would begin to speak about gravity itself.

\section*{Acknowledgments}
We thank Eliahu Cohen for pointing out the role of the energy variance, which led to the analysis of Section~\ref{sec:memory}, for questioning the thermodynamic argument, and for the observation, made at the San Diego meeting on time travel, that the spatial super-translation of \cite{AharonovCarmi2018} should have a temporal counterpart with clocks; Nadav Shnerb for comments; and Yarin Mimon for building the platform on which this collaboration runs, and for much else. Equations and part of the analysis were developed with the assistance of an AI system.

\end{document}